\documentclass[11pt]{article}
\usepackage{fancyhdr}
\usepackage{isomath}
\usepackage{amsmath}
\usepackage{amsbsy}
\usepackage{amssymb}
\usepackage{amscd}
\usepackage{amsfonts}
\usepackage{graphicx,color}
\usepackage{verbatim}
\usepackage{euscript}
\usepackage{alltt}
\usepackage{stmaryrd}
\usepackage{subfigure}
\usepackage{relsize}

\DeclareGraphicsExtensions{.eps,.pdf}
\usepackage{amsmath}
\usepackage{amsbsy}
\usepackage{amssymb}
\usepackage{amscd}
\usepackage{amsfonts}

\newcommand{\bfb}{{\mathbold b}}
\newcommand{\bfc}{{\mathbold c}}

\newcommand{\bfe}{{\mathbold e}}
\newcommand{\bff}{{\mathbold f}}

\newcommand{\bfn}{{\mathbold n}}

\newcommand{\bft}{{\mathbold t}}
\newcommand{\bfu}{{\mathbold u}}
\newcommand{\bfv}{{\mathbold v}}

\newcommand{\bfx}{{\mathbold x}}

\newcommand{\bfA}{{\mathbold A}}

\newcommand{\bfI}{{\mathbold I}}

\newcommand{\bfM}{{\mathbold M}}

\newcommand{\bfT}{{\mathbold T}}

\newcommand{\bfV}{{\mathbold V}}

\newcommand{\beq}{\begin{equation}}
\newcommand{\eeq}{\end{equation}}
\newcommand{\beqs}{\begin{eqnarray}}
\newcommand{\eeqs}{\end{eqnarray}}
\newcommand{\beql}{\begin{equation} \label}

\newcommand{\del}{\partial}

\newcommand{\bfveps}{\mathbold{\varepsilon}}

\newcommand{\bfzero}{\mathbf{0}}

\newcommand{\deriv}[2]{\frac{d #1}{d #2}}

\newcommand{\veps}{\varepsilon}

\usepackage[margin=1in]{geometry}

\date{}
\begin{document}
\title{A possible link between brittle and ductile failure by viewing fracture as a topological defect}

\author{Amit Acharya\thanks{Department of Civil \& Environmental Engineering, and Center for Nonlinear Analysis, Carnegie Mellon University, Pittsburgh, PA 15213, email: acharyaamit@cmu.edu.}}
\maketitle

\begin{abstract}
\noindent A continuum model of fracture that describes, in principle, the propagation and interaction of arbitrary distributions of cracks and voids with evolving topology without a ‘fracture criterion’ is developed. It involves a ‘law of
motion’ for crack-tips, primarily as a kinematical consequence coupled with thermodynamics. Fundamental kinematics endows the crack-tip with a topological
charge. This allows the association of a kinematical conservation law for the charge,
resulting in a fundamental evolution equation for the crack-tip field, and in turn the crack field.
The vectorial crack field degrades the elastic modulus in a physically justified anisotropic manner. The mathematical structure of this conservation law allows an additive ‘free’ gradient of a scalar field in the evolution of the crack field. We associate this naturally emerging scalar field with the porosity that arises in the modeling
of ductile failure. Thus, porosity-rate gradients affect the evolution
of the crack-field which, then, naturally degrades the elastic modulus, and it is through this fundamental mechanism that spatial gradients in porosity growth affects the strain-energy density
and stress carrying capacity of the material - and, as a dimensional consequence related to fundamental kinematics, introduces a length-scale in the model. The key hypothesis of this work is that brittle fracture is energy-driven while ductile fracture is stress-driven; under overall shear loadings where mean stress vanishes or is compressive, shear strain energy can still drive shear fracture in ductile materials.
\end{abstract}

\section{Introduction}
 Fracture of brittle and ductile materials is the most common mode of final failure in solids. Fracture is observed to occur in varying forms - from a single macroscopic crack propagating from a pre-existing notch and a single crack branching into daughter cracks,  to a distribution of smaller cracks forming an evolving swarm. Fracture in brittle materials, e.g. high-strength, low weight ceramics \cite{clayton2015nonlinear,misseroni2016experiments}, or brittle fracture in metals, e.g., HCP and BCC metals \cite{pineau2016failure}, is observed to occur along sharp, well-defined cleavage planes whereas ductile fracture (e.g. in structural metals) occurs by the nucleation, growth, and coalescence of voids \cite{teirlinck1988fracture,weck20062d,mcclintock1966ductile}. In general, tensile hydrostatic stress states promote fracture but fracture in ductile (and brittle) materials has been observed under imposed shear loading (with no hydrostatic component). Finally, fracture occurs under quasi-static to highly dynamic loading scenarios.
The goal of this note is to explore possible connections between the modeling of brittle and ductile fracture based on fundamental kinematical and continuum mechanical grounds.

\section{The mathematical model}
In this section we briefly reproduce some basic material from \cite{acharya2018fracture} to set the stage for its extension for coupling brittle cracking to the ductile fracture mechanism of void growth. 
\subsection{Kinematic descriptors of fracture and their physical motivation}\label{sec:kin_frac}
\begin{figure}[h]
\centering
\includegraphics[width=6.5in, height=2.7in]{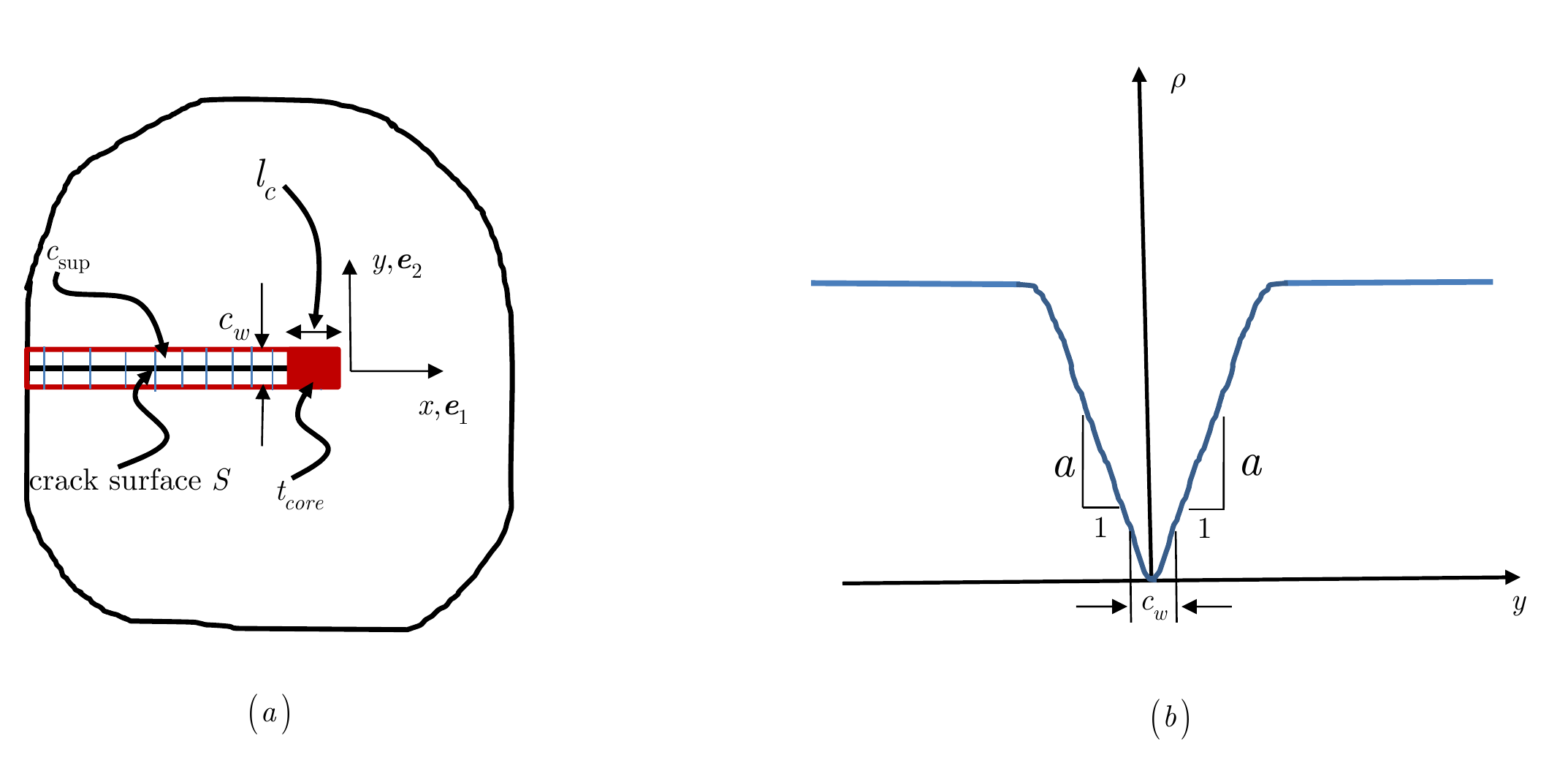}
\caption{{\sf Geometry of an idealized crack}.}
\label{fig:crack_geom}
\end{figure}
 With reference to Figure \ref{fig:crack_geom}, consider the following situation: we consider the region $c_{\sup}$ as divided into a set of disjoint `vertical' neighborhoods as shown by the blue lines, each centered around a point $\bfx_S \in S$. We refer to each such neighborhood as $N(\bfx_S)$. We think of measuring the mass density field $\tilde{\rho}$ around $\bfx_S$ and call it the local mass density field in $N(\bfx_S)$. We assume this locally measured mass density field to be continuous in the neighborhood (possibly taking on the value of 0 at some points). At the scale of observation, let it go to $0$ on the crack surface $S$. Assume the locally measured density variation at each $\bfx_S$ along the normal direction to the surface $S$ be of the form as shown in Fig. \ref{fig:crack_geom}(b). The macroscopic crack surface of $0$-thickness actually is spread over the region $c_{\sup}$ where the density may (or may not) be smooth/differentiable; but we assume that we are unable to resolve the variation of the measured density gradient in $c_{\sup}$. Thus, $grad \tilde{\rho}$ appears discontinuous across the crack surface. However, since $\tilde{\rho}$ is continuous
\begin{equation}\label{eqn:density_jump}
\llbracket grad \tilde{\rho} \rrbracket = \varphi \bfn \ \ \ \mbox{necessarily,}
\end{equation}
where $\bfn$ is an arbitrarily chosen orientation for the crack surface, $\llbracket \cdot \rrbracket$ denotes a jump, and $\varphi$ is a scalar field on $ S \cap N(\bfx_S)$, i.e. $grad \tilde{\rho}$ can jump only in the normal direction to the surface.

Assume that we are able to choose the orientation field $\bfn$ for the crack surfaces in the body at any given time in a continuous way except possibly at points where $\varphi = 0$. We now define the \textit{crack field} as
\begin{equation*}
\bfc(\bfx) := \begin{cases}
\frac{\varphi(\bfx_S) \bfn(\bfx_S)}{c_w} \ \ \mbox{for} \  \bfx \in N(\bfx_S) \in c_{\sup} \\
0 \ \ \ \mbox{for} \ \bfx \ \mbox{outside} \  c_{\sup},
\end{cases}
\end{equation*}
where $c_w$ is the width of the region $c_{sup}$ measured along the direction $\bfn$, pointwise, and the \textit{crack tip} field as
\begin{equation}\label{eqn:t_c}
\bft := -curl \bfc, \ \ \ \mbox{globally on the body}
\end{equation}
(with the minus sign for convenience). Assuming $\bfn$ to be oriented in the direction $\bfe_2$ in Fig. \ref{fig:crack_geom}, $\varphi(\bfx) = 2a$ for $\bfx \in c_{\sup}\backslash t_{core}$. Thus, by thinking of the jump of the local density-gradient to be spread out over the layer, we may interpret the field $\bfc$ as an approximation to the directional derivative of the local density gradient field in the direction $\bfn$.

With the above argument as physical motivation, we now consider $\bfc$ and $\bft$ as continuous fields for the sake of developing the mechanical model (as customary in mechanics). To justify the terminology for the crack tip field, with reference to Fig. \ref{fig:crack_geom}, let the density gradient go from some constant value in $c_{\sup}$, say $\varphi_0$ ($\varphi_0 = 2a$ in the example considered in Fig. \ref{fig:crack_geom}), to $0$ over the length of the region $t_{core}$. Let $\bfc$ vary in-plane for simplicity. Then
\begin{equation*}
\begin{split}
& \bfc = c_2\, \bfe_2\\
& -curl \bfc = \bft = - e_{312}\, c_{2,1} \,\bfe_3 = \frac{\varphi_0 - 0}{l_c \cdot c_w} \bfe_3,
\end{split}
\end{equation*}
and this is non-vanishing only in the region $t_{core}$. Thus $curl \bfc$ identifies the crack-tip region. It is also important to note that the $curl$ is insensitive to the large gradients in $\bfc$ in the vertical direction across the horizontal boundaries of the layer. 

\subsection{Governing field equations}
Keeping crack extension through crack-tip motion in mind, we note first that
\begin{equation*}
\bft = -curl \bfc \implies \dot{\bft} = -curl \dot{\bfc}
\end{equation*}
and $\dot{\bfc}$ should be a function of the crack-tip field $\bft$ and the crack-tip velocity (with respect to the material) $\bfV$, postulated to be a field in this model. The crack-tip is identified by the field $\bft$ and keeping within the confines of `local' and simplest theory, it is natural to look for a relation of the type $\dot{\bfc} = \bff (\bft, \bfV)$. It is established in \cite{acharya2018fracture} on fundamental grounds that the crack-tip field carries a topological charge and that its evolution is governed by a conservation law for the charge given by
\begin{equation}\label{eqn:loc_conserv_law}
\dot{\bft} = - curl (\bft \times \bfV).
\end{equation}
Equation (\ref{eqn:loc_conserv_law}) implies an evolution equation for the crack field $\bfc$ of the form
\begin{equation}\label{c_evol}
\dot{\bfc} =  -curl \bfc \times \bfV
\end{equation}
up to a `free' gradient. If the crack field $\bfc$ is restricted to evolve only by motion of the crack-tip field $\bft$ as in brittle fracture, then this gradient vanishes. However, in modeling ductile fracture, \textit{linking this to the gradient of the porosity growth field, the latter typically the fundamental ingredient of all ductile fracture models (e.g. GTN), demonstrates a physical mechanism (as opposed to ad-hoc modification) for such porosity growth to couple to the degradation of elastic moduli through its effect on the growth of the crack field}. We note that our model is discerning enough to not allow a density of voids (a volumetric density of objects concentrated on points in the singular limit) to instantaneously produce a crack-tip (an areal density of objects concentrated on curves in the singular limit), objects that produce very different stress concentrations under applied stress. Furthermore, the model also implies that as long as crack-tips are not inserted from outside the whole body under consideration, the crack-tips nucleated must be closed loops (encircling (non)planar penny-shaped regions, e.g.) or ensure that the total topological charge within the whole body does not evolve in time.

We now detail this physical coupling of `brittle' cracking to `ductile' porosity evolution - denoted by a scalar field $f$, purely on the basis of kinematics in the first instance, and then develop its thermodynamic consequences. We do so in the context of the GTN model, as expounded on in \cite{nahshon2008modification, benzerga2016ductile}. Based on what has been said above,  \eqref{c_evol} is modified to
\begin{equation}\label{eqn:c_evol_mod}
\dot{\bfc} = -curl \bfc \times \bfV + m \ grad \dot{f},
\end{equation}
where $m$ is a mobility constant with physical dimensions of $\frac{Mass \  density}{Length}$, required on dimensional grounds (note that based on its physical meaning, $\bfc$ has physical dimensions of $\frac{Mass \ density}{Length^2}$ and the porosity, $f$, is, of course, dimensionless). With reference to Fig. \ref{fig:crack_geom}, \textit{in all that follows, we will use the normalization $\widetilde{\bfc} = \frac{c_w l_0}{2 \rho_M}\bfc$, where $\rho_M$ is the mass density of the intact matrix and $l_0 = \frac{\rho_M}{a}$; we will also define a single length scale $l := \sqrt{\frac{c_w l_0}{2}}$ and drop the overhead $\  \widetilde{} \ $ on $\widetilde{\bfc}$ for convenience}. With this understanding, the evolution equation for the \textit{normalized crack field} becomes
\begin{equation}\label{eqn:norm_c_evol}
\dot{\bfc} = - curl \bfc \times \bfV + \mathfrak{p} \, grad \dot{f}; \qquad \mathfrak{p} := \frac{m l^2}{\rho_M}.
\end{equation}
In this model of coupled ductile-brittle damage, the porosity $f$ appears as a fundamental kinematic field as well.

In this preliminary note a `geometrically linear,' or small deformation theory is considered which, nevertheless, is materially nonlinear. There are good reasons, based on our past experience with theories of similar type \cite{garg2015study}, to expect crack nucleation, in what would be the purely brittle case in the present context, to require geometric nonlinearity - we defer this for future work, especially since ductile damage nucleation can be incorporated (phenomenologically) through the evolution equation for the porosity as in the GTN model. The governing field equations for the model are
\begin{equation}\label{eqn:gov_eqn}
\begin{split}
& \rho_0 \ddot{\bfu} = div \, \bfT + \bfb\\
& \dot{\bfc} = - curl \bfc \times \bfV + grad \, s,
\end{split}
\end{equation}
where $\rho_0$ is the time-independent mass density field corresponding to the reference configuration of the body from which all displacements are measured, $\bfT$ is the symmetric stress tensor, $\bfb$ is the body force density per unit volume of the reference configuration, $\bfu$ is the displacement field,  $\dot{\bfu} = \bfv$ is the material velocity field, and we consider the `source' $s = \mathfrak{p} \dot{f}$. Also, all differential operators $div$, $curl$ are written with respect to the fixed reference configuration.

\subsubsection{Reversible response functions and driving forces for dissipation}
We consider mechanical effects only. Assume a free-energy density function (per unit volume of reference configuration) given as
\[
\psi(\bfveps^e, \bfc,\bft)
\]
where $\bfveps^e = \bfveps - \bfveps^p$ and $\bfveps = sym(grad\bfu)$, with $\bfu$ being the displacement field, and $\bfveps^p$ is the symmetric plastic strain tensor.

The mechanical dissipation is defined as the power supplied by the external forces (tractions and body forces) less the rate of change of kinetic energy and the power stored in the body:
\begin{equation}\label{dissip_1}
\mathsf{D} = \int_{\partial V} \bft \cdot \dot{\bfu} \,da + \int_V \bfb \cdot \dot{\bfu}\, dv - \deriv{}{t} \int_V \psi \, dv - \deriv{}{t} \int_V \frac{1}{2} \rho_0 |\dot{\bfu}|^2 \,dv.
\end{equation}
Using the governing equations (\ref{eqn:gov_eqn}), the dissipation can be expressed as
\begin{equation}\label{dissip_2}
\begin{split}
\mathsf{D}  = &\int_V \left( \bfT - \partial_{\bfveps^e} \psi \right) : grad\bfv \,dv  \\
& + \int_V \del_{\bfveps^e} \psi : \dot{\bfveps}^p \, dv + \int_V \left\{   \left[  - \partial_{\bfc} \psi + curl\, \partial_{\bft} \psi \right] \times \bft \right\} \cdot \bfV \, dv  + \int_V div( \mathfrak{p}\, \partial_\bfc \psi) \dot{f} \, dv \\
& + \int_{\partial V} \bfV \cdot \left[ \left( \partial_{\bft} \psi \times \bfn \right) \times \bft \right] \, da - \int_{\del V} \mathfrak{p} \, (\partial_\bfc \psi \cdot \bfn) \, \dot{f} \, da.
\end{split}
\end{equation}
Following the GTN \cite{benzerga2016ductile, nahshon2008modification} model, we now assume that the porosity evolution takes the form
\begin{equation}\label{porosity}
\dot{f} = (1 - f) tr(\dot{\bfveps^p}),
\end{equation}
(where we have not included a nucleation term for ease of exposition). On demanding classical hyperelasticity be recovered in the absence of plasticity and crack evolution and porosity growth, we obtain the stress relation
\begin{equation}\label{eqn:stress}
\bfT = \partial_{\bfveps^e} \psi
\end{equation}
and note the driving forces in the bulk for the mechanisms of plasticity and the crack-tip advance as
\begin{equation}\label{bulk_driv_forces}
\begin{split}
\dot{\bfveps}^p & \leadsto \bfT + (1 - f) div( \mathfrak{p}\,\partial_\bfc \psi) \bfI \\
\bfV & \leadsto \left[  - \partial_{\bfc} \psi + curl\, \partial_{\bft} \psi \right] \times \bft.
\end{split}
\end{equation}
Driving forces at the boundary also emerge as
\[
\bfV \leadsto \left( \partial_{\bft} \psi \times \bfn \right) \times \bft; \qquad \qquad {\dot{\bfveps}^p} \leadsto  (1 - f) (\mathfrak{p}\,\partial_\bfc \psi \cdot \bfn)  \bfI.
\]

\subsubsection{\emph{Proposed nonlocal, modified GTN model of coupled brittle-ductile damage}}
~Ignoring the boundary dissipation terms for simplicity and motivated by the form of the bulk driving force for $\dot{\bfveps}^p$, the closed, governing equations of the proposed nonlocal, modified GTN model become \cite{benzerga2016ductile,nahshon2008modification}
\begin{subequations}\label{eqn:gov_eqn_mod_GTN}
\begin{align}
&\psi  = \hat{\psi}(\bfveps^e, {\color{blue} \bfc, curl\bfc})\\
& \quad \sigma_{eq} = \sqrt{\frac{3}{2} \bfT':\bfT'} \ ; \qquad  \sigma_m = \frac{1}{3} tr(\bfT) \ ; \qquad {\color{blue} \sigma_m^* = \sigma_m + (1 - f) div(\mathfrak{p}\,\del_\bfc \psi)}  \label{eqn:sigma_m*} \\
& \Phi(\bfT, f^{*}(f), {\color{blue} div\,(\mathfrak{p} \del_\bfc \psi)}) = \frac{\sigma_{eq}^2}{\overline{\sigma}^2} + 2 q_1 f^{*} \cosh{\frac{3 q_2 {\color{blue} \sigma_m^{*}}}{2 \overline{\sigma}}} - (1 + q_3 f^{*2})\\ 
& f^*(f) = 
\begin{cases}
f  & \mbox{if} \  f < f_c\\
f_c + \frac{\left(\frac{1}{q_1} - f_c \right) (f - f_c)}{f_f - f_c}  & \mbox{if} \  f \geq f_c 
\end{cases}\\
& \del_\bfT \Phi = \frac{3 \bfT'}{\overline{\sigma}^2} + \frac{f^* q_1 q_2}{\overline{\sigma}}\sinh\frac{3 q_2 {\color{blue} \sigma_m^{*}}}{2 \overline{\sigma}} \bfI\\
& \dot{f}  = (1 - f)tr(\dot{\bfveps}^p) = (1 - f) \frac{3 \Lambda f^* q_1 q_2}{\overline{\sigma}} \sinh\frac{3 q_2 \left[ \sigma_m + {\color{blue} (1 - f) div \,  (\mathfrak{p} \del_\bfc \psi) }\right]} {2 \overline{\sigma}} \label{eqn:fdot}\\
& \dot{\bfveps}^p  = \Lambda \, \partial_{\bfT} \Phi; \quad \Lambda \Phi = 0; \quad \Phi \leq 0; \quad \Lambda \geq 0 \label{13_g}\\
& \overline{\sigma} = \overline{\sigma}(\overline{\veps}) \qquad \mbox{ given stress-plastic strain curve in uniaxial tension for matrix material}\\
& \dot{\overline{\veps}} = \frac{\Lambda (1 - f) \bfT : \del_\bfT \Phi}{\overline{\sigma}} \label{13_i}\\
&\rho_0 \ddot{\bfu}  = div\, \bfT + \bfb \label{13_j}\\
&{\color{blue} \nonumber
\bfA = curl\,\bfc \times \bfM \left[ \left\{ - \partial_{\bfc} \hat{\psi} - curl \left(\partial_{curl \bfc} \hat{\psi} \right) \right\} \times curl \bfc \right] + \mathfrak{p} \, grad \left[ (1 - f) tr(\dot{\bfveps}^p) \right]} \\
& {\color{blue} \bfA = curl\,\bfc \times \bfM \left[ \left\{ - \partial_{\bfc} \hat{\psi} - curl \left(\partial_{curl \bfc} \hat{\psi} \right) \right\} \times curl \bfc \right] }  \label{eqn:A} \\
&  {\color{blue} \qquad +  \  \mathfrak{p} \, grad \left[ (1 - f) \frac{3 \Lambda f^* q_1 q_2}{\overline{\sigma}} \sinh\frac{3 q_2 \left[ \sigma_m + (1 - f) div \,  (\mathfrak{p} \del_\bfc \psi) \right]} {2 \overline{\sigma}} \right]} \nonumber \\
&{\color{blue} \dot{\bfc}  = 
\begin{cases}
\bfA \ \ \mbox{if} \ \ \frac{\bfc}{|\bfc|}\cdot \bfA > 0\\
\bfzero \ \ \mbox{otherwise}
\end{cases} \mbox{irreversibility of cracking induced damage} \Leftrightarrow \dot{|\bfc|} \geq 0}, \label{eqn:cdot}
\end{align}
\end{subequations}
where $\bfveps^p, f, \overline{\veps}, {\color{blue} \bfc}$ are the state variables that need to be evolved, and the terms marked in blue are the proposed modifications to the GTN model (we recall that $\bfc$ is vector-valued). In the above, $\bfM$ is a symmetric, positive definite tensor of crack mobility that could take the isotropic form $\bfM = \frac{1}{B} \bfI$, where $B > 0 $ is a scalar drag coefficient, $\mathfrak{p}$ is the mobility scalar discussed earlier, $\bfT'$ is the stress deviator, $q_1, q_2, q_3, f_c, f_f$ are specified parameters of the GTN model, and we have assumed the plasticity to be rate-independent (but nevertheless the overall model is generally rate-dependent due to the first term in the expression for $\bfA$). It can be checked that the above nonlocal, modified, GTN model results in non-negative dissipation.

A typical candidate for the energy density would be
\begin{equation} \label{fe_density}
\hat{\psi}(\bfveps^e, \bfc, curl \, \bfc) = \psi_E(\bfveps^e, \bfc) + \eta(|\bfc|) + \mathfrak{t} |curl\bfc|^2,
\end{equation}
where $\psi_E$ represents the elastic strain energy density of the material with its elastic modulus degraded to reflect damage due to cracking represented by $\bfc$ but at the same time providing resistance to interpenetration of crack-flanks, $\eta$ is a non-convex function representing an energy barrier to damage from an undamaged state, and $\mathfrak{t}$ is a small parameter regularizing the crack-tip (but not the crack layer). The term $|curl\,\bfc|^2$ may be considered as the lowest-integer-order approximation of any smooth function that assigns an energy cost to the formation of a crack-tip, the latter kinematically characterized by a non-vanishing $curl \, \bfc$. Next, we describe the modeling of $\psi_E$ and $\eta$.

\subsection{Elastic strain energy density of cracked material preventing interpenetration of crack flanks}
Let $H(x) = 0$ for $x \leq 0$ and $H(x) = 1$ for $x > 1$ be the Heaviside step function. For ease of exposition, we assume the intact matrix material to be elastically isotropic with the $4^{th}$-order tensor of elastic moduli given by $\mathbb{C} = \lambda \bfI \otimes \bfI + 2\mu \mathbb{I}$, where $\lambda, \mu$ are the Lame parameters and $\mathbb{I}$ is the identity tensor on the space of symmetric second-order tensors. We assume the $\widetilde{\lambda}(|\bfc|), \widetilde{\mu}(|\bfc|)$ are two functions on the space of non-negative scalars representing monotonically decreasing degradation functions for elastic moduli as a function of magnitude of cracking (our model does not provide guidance on these choices, just as in phase-field models \cite{bourdin2000numerical, borden2012phase}, apart from requiring them to be convex; there does exist an extensive literature based on homogenization to estimate such effects due to cracking). Define $\widetilde{\mathbb{C}} = \widetilde{\lambda}(|\bfc|) \bfI \otimes \bfI + 2 \widetilde{\mu}(|\bfc|) \mathbb{I}$. Further define $\widehat{\bfc} = \frac{\bfc}{|\bfc|}$, $\veps^e_c = \widehat{\bfc} \cdot \bfveps^e \widehat{\bfc}$, and $\bfveps^e_\perp = \bfveps^e - \veps^e_c \, \widehat{\bfc} \otimes \widehat{\bfc}$. Then
\begin{align}\label{eqn:se_density}\nonumber
2 \psi_E(\bfc, \bfveps^e) & = H(|\bfc|)\left[ H(\veps^e_c) \bfveps^e:\widetilde{\mathbb{C}} \bfveps^e + (1 - H(\veps^e_c)) \left\{ \bfveps^e_\perp: \widetilde{\mathbb{C}} \bfveps^e_\perp +  \veps^{e2}_c \underline{(\widehat{\bfc} \otimes \widehat{\bfc}): \mathbb{C} (\widehat{\bfc} \otimes \widehat{\bfc})} \right\} \right]\\
&\quad + (1 - H(|\bfc|)) \bfveps^e : \mathbb{C} \bfveps^e.
\end{align}
In the above, the underlined term can as well be replaced by a contact `stiffness' separate from the material elasticity if so desired. The physical ideas embodied in \eqref{eqn:se_density} are as follows: for any material point that is considered as cracked
\begin{itemize}
\item if the elastic strain component in the direction normal to the local crack is extensional, then the elastic response is damaged for all strain modes;
\item if the elastic strain component along the local crack normal direction vanishes or is compressional, then all strain modes except the one along $\widehat{\bfc} \otimes \widehat{\bfc}$ respond in a damaged manner whereas along the crack normal direction the compression is resisted as if the material was undamaged, or according to some prescribed contact stiffness (cf. \cite{tvergaard2009behaviour}) (either way, crack-flank interpenetration is resisted);
\item if the material point is uncracked, then the elastic response of the material is that of an intact material.
\end{itemize}

\subsection{Crack energy barrier density} 
\begin{figure}[h]
\centering
\includegraphics[width=6.5in, height=2.5in]{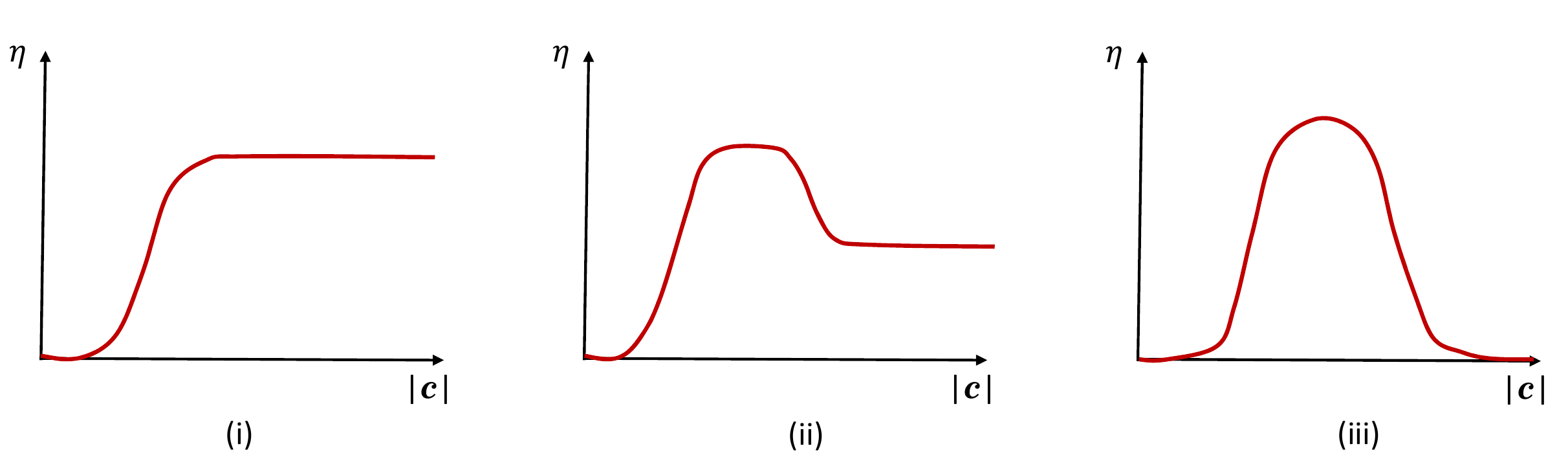}
\caption{{\sf Qualitative, local crack energy density functions}.}
\label{fig:crack_eta}
\end{figure}

The energy density function $\eta$ represents the energy cost incurred at a material point due to cracking. Fig. \!\ref{fig:crack_eta} represents three different possibilities, corresponding to i) Griffith-type (local) surface energy barrier where the energy cost as a function of cracking intensity stabilizes, ii) where the  local energy barrier decreases beyond its maximum with increased cracking and then stabilizes, and iii) where the local energy barrier decreases to zero beyond its maximum with increased cracking intensity. The first two functions correspond to models of some surface energy being assigned to fully cracked neighborhoods, whereas the last one reflects all elastic energy of cracking being dissipated. In purely Mode I situations, the last option has the possibility of predicting irreversibility of cracking without any added modeling, e.g. as in \eqref{eqn:cdot}.

\subsubsection{Discussion}
 We make the following observations regarding the salient characteritics of the model:
\begin{enumerate}
    \item The cracked elastic energy density \eqref{eqn:se_density} we propose is finely and directly adapted, in a physically transparent manner, to the modeling of resistance to interpenetration of crack-flanks under compression normal to the crack surface. This is enabled by the fact that the crack field is a naturally emergent vector field normal to the local crack surface, as opposed to a scalar damage field of ambiguous physical origin in phase field models \cite{bourdin2000numerical,hakim2009laws,miehe2010phase, amor2009regularized,borden2012phase,tu2020coupled}. In all of these cited works related to phase-field models, principal values and directions of the strain tensor or the hydrostatic part of the strain tensor are used to approximately achieve the stated goal of ``the intent of both models is similar, that is, to maintain resistance in compression and, in particular, during crack closure,'' which is a quote from the paper of \cite{borden2012phase} in comparing and contrasting this aspect of crack mechanics in their model (following \cite{miehe2010phase}) with \cite{amor2009regularized}.
    \item It is a well-established experimental fact that shear fracture occurs in ductile specimens under $0$ or negative mean stress \cite{mcclintock1971plasticity,johnson_cook,bao2004fracture, barsoum2007rupture,nahshon2008modification}. The most prominent models for modeling ductile fracture - the Gurson model (later improved to be the GTN model) and the Johnson-Cook model - emphasize the role of increasing stress triaxiality $\left( \frac{\sigma_m}{\sigma_{eq}} \right)$ in reducing fracture strains of ductile materials, with predictions of no fracture for $0$ mean stress, which is not consistent with observations. The fundamental mechanism of shear fracture of ductile materials, based on the work of McClintock \cite{mcclintock1971plasticity} and Teirlinck et al. \cite{teirlinck1988fracture}, is stated by Nahshon and Hutchinson \cite{nahshon2008modification} to be ``void-sheet formation as the underlying mechanism wherein it is supposed that under shearing voids increase their effective collective cross-sectional area parallel to the localization band without an accompanying increase in void volume. Localization in shear in micro bands linking voids is evident in the model voided materials tested by Weck et al. \cite{weck20062d}.''
    
    With this backdrop, we note that in our model the `brittle' cracking mechanism arising from $\bfc$ affecting the elastic modulus results in energy-driven as opposed to stress-driven fracture. This coupled with the modification to the mean stress $\sigma_m$ embodied in $\sigma_m^*$ \eqref{eqn:sigma_m*} implies that even under $0$ mean stress, there is a driving force for the evolution of porosity \eqref{eqn:fdot} as well as the evolution of $\bfc$ \eqref{eqn:A}-\eqref{eqn:cdot}. Moreover, based on what has been described in Sec. \ref{sec:kin_frac}, the evolution of $\bfc$ is very much adapted to ``under shearing voids increase their effective collective cross-sectional area parallel to the localization band without an accompanying increasing in void volume.'' Thus, the proposed model needs to be explored in-depth to examine its potential for describing shear fracture of ductile materials under vanishing or compressional mean stress.
    \item It is generally believed that the effect of porosity on elastic modulus is small and such effects are neglected. However, such small effects of over small length scales can have order 1 effects in spatial gradients, which is the essential modification in the proposed model that arises from the presence of $div (\mathfrak{p} \, \del_\bfc \psi)$ in $\sigma_m^*$. As already mentioned, such effects then permeate into the evolution of porosity $f$ and the crack-field $\bfc$. A physical way to see this is that a through-crack in a body contributes very little to volumetric damage - indeed, an idealized crack represented by a 2-d surface contributes to no volumetric porosity - however, it results in complete loss of stress-carrying capacity. Thus, the effect of cracking on elastic modulus is a different physical mechanism than the effect of porosity on elastic modulus degradation; indeed, it is porosity gradients that affect cracking in this model.
    
    In this regard, we note recent work \cite{dorhmi2020homogenization} that shows the effect of elastic modulus degradation due to porosity, as well as well-established ideas and methods (cf. \cite{castaneda1997nonlinear}) to estimate elastic strength degradation due to porous microstructures. An additional dependence of the elastic strain energy density function on $f$ is easily accommodated in the present formalism and will result in an additional driving force contribution in $\sigma_m^*$ \eqref{eqn:sigma_m*}.
    \item \label{M=0} When the mobilities $\bfM = \bfzero$ and $\mathfrak{p} = 0$, \eqref{eqn:gov_eqn_mod_GTN} reduces to the GTN model. For $\bfM = \bfzero$, one still has a thermodynamically consistent `nonlocal' generalization of the Gurson model. Eqns. \eqref{eqn:A} and \eqref{eqn:cdot} imply that $\bfc$ has to be a gradient of a scalar field and, along with \eqref{eqn:fdot}, one obtains that this scalar is the porosity $f$ (up to a spatially constant function of time, which we assume to vanish). Thus $\bfc \equiv grad\, f$ in this idealization, and in this damage physics related only to volumetric porosity, we assume $\eta \equiv 0$ and $\mathfrak{t} \equiv 0$. It is instructive at this point to consider an expansion of \eqref{eqn:fdot} for small $\mathfrak{p} > 0$ about $0$:
    \begin{align}
    \dot{f} & = (1-f) \frac{3 \Lambda f^* q_1 q_2}{\overline{\sigma}} \left(\sinh\frac{3 q_2 \sigma_m}{2 \overline{\sigma}} + \cosh \frac{3 q_2 \sigma_m}{2 \overline{\sigma}} (1-f) div \, \del_\bfc \psi \ \mathfrak{p} + \mathcal{O}(\mathfrak{p}^2) \right) \nonumber\\
    & = (1-f) \frac{3 \Lambda f^* q_1 q_2}{\overline{\sigma}} \left( \sinh\frac{3 q_2 \sigma_m}{2 \overline{\sigma}} + \mathfrak{p} \, (1-f) \cosh \frac{3 q_2 \sigma_m}{2 \overline{\sigma}} \left[ \del_{\bfveps^e \bfc} \psi \  \vdots \  grad \, \bfveps^e \right] \right) \nonumber\\
     & \quad + \mathfrak{p} \, (1-f)^2 \ \frac{3 \Lambda f^* q_1 q_2}{\overline{\sigma}} \cosh \frac{3 q_2 \sigma_m}{2 \overline{\sigma}} \left[  \del_{\bfc \bfc} \psi : grad^2 f \right] \label{eqn:porosity_diffusion},
    \end{align}
    on formally ignoring the $\mathcal{O}(\mathfrak{p}^2)$ terms. The last line of \eqref{eqn:porosity_diffusion} is particularly illuminating - with $\mathfrak{p} > 0$ and $\del_{\bfc \bfc} \psi$ assumed convex, this is a \textit{completely defined diffusive regularization to the GTN porosity evolution, with no adjustable parameters once a commitment to the physically realistic elastic energy density function $\psi_E$ \eqref{eqn:se_density} has been made} (we note that the convexity requirement allows degradation of elastic modulus as a function of $|\bfc|$). A particularly familiar simplification is if $\del_{\bfc \bfc} \psi$ were to be a positive scalar multiple of the second order Identity tensor, in which case one recovers the phenomenologically introduced Laplacian regularization for porosity damage \cite{ramaswamy11998finite}; in the present rate-independent model, both the yield function and the evolution equations have gradient terms in them even in the `purely ductile' setting ($\bfM =\bfzero$). Finite element based computational methods for such situations, even at finite deformations, are available \cite{ramaswamy11998finite,ramaswamy21998finite}. For a micromechanics-based gradient regularization of damage due to void growth see \cite{gologanu1997recent}.
    \item The overall physical mechanism implied by the proposed model can be summarized as follows: Remark \ref{M=0} lays bare the role that the proposed theory brings to the physical regularization of porosity-induced ductile damage. As is well-understood by now, due to the softening in material strength produced by damage, elastic unloading takes place outside of localizing 3-d damage zones, with these zones decreasing in width to a vanishing thickness around 2-d surfaces/regions in the limit - the gradient regularization produces a damage zone of finite width in the transverse direction to the thin zones. However, this mechanism by itself does not suggest anything about the \textit{longitudinal propagation} of such 2-d thin regions, as would be required by the void linking mechanisms of \cite{mcclintock1971plasticity,tvergaard2009behaviour,nielsen2012collapse,morin2016application} (we note that the elastic strain gradient term does  provide a fundamental and interesting nucleation mechanism, whose role needs to be explored). In the proposed model, once the rate of porosity gradients approach $\sqrt{\mathfrak{p}}$ in magnitude, they start to affect the development of $\bfc$ and once $\bfc$ is generated, the evolution of this field (for $\bfM \neq \bfzero$) occurs primarily through the lateral expansion of these thin damage zones by the motion of the `crack tips,' or the terminating boundary of these thin zones.
    \item The proposed model coupling `brittle' crack growth and decohesion to ductile damage through void growth provides a fundamental basis for extending phase-field like models for the modeling of ductile fracture, the state-of-the-art of which can be seen in \cite{miehe2016phase} and \cite{ambati2015phase}, the latter involving ad-hoc nonlinear modification of the phase field variable by the equivalent plastic strain while demonstrating encouraging results. We mention here the trend towards more physical representation of ductile fracture in phase-field modeling in the very recent interesting work of \cite{tu2020coupled}.
    
    Nevertheless, much further work is necessary to understand the full implications of the presented model and to compare and contrast its predictions to what is currently known.
\end{enumerate}
\section*{Acknowledgment}
It is a pleasure to acknowledge and thank Leo Morin for insightful comments and discussion.

\bibliographystyle{ieeetr}
\bibliography{brittle_ductile_ref}

\end{document}